# Investigation of the interfacial Dzyaloshinskii-Moriya interaction sign in Ir/Co$_2$FeAl systems by Brillouin light scattering

M. Belmeguenai[1,*], M. S. Gabor[2,**], Y. Roussigné[1], T. Petrisor jr[2], R.B. Mos[2], A. Stashkevich[1,3], S. M. Chérif[1] and C. Tiusan[2,4]

[1]*LSPM, CNRS-Université Paris 13, Sorbonne Paris Cité, 99 avenue Jean-Baptiste Clément Université Paris 13, 93430 Villetaneuse, France*
[2]*Center for Superconductivity, Spintronics and Surface Science, Technical University of Cluj-Napoca, Str. Memorandumului No. 28 RO-400114 Cluj-Napoca, ROMANIA*
[3]*International laboratory MultiferrLab, ITMO University, St. Petersburg 197101, Russia*
[4]*Institut Jean Lamour, CNRS, Université de Nancy, BP 70239, F–54506 Vandœuvre, France*

***Abstract-***Co$_2$FeAl (CFA) ultrathin films, of various thicknesses (0.9 nm≤$t_{CFA}$≤1.8 nm), have been grown by sputtering on Si substrates, using Ir as a buffer layer. The magnetic properties of the structures have been studied by vibrating sample magnetometry (VSM), miscrostrip ferromagnetic resonance (MS-FMR) and Brillouin light scattering (BLS) in the Damon-Eshbach geometry. VSM characterizations show that films are mostly in-plane magnetized and the perpendicular saturating field increases with decreasing CFA thickness suggesting the existence of interface anisotropy. The presence of magnetic dead layers of 0.44 nm has been detected by VSM. The MS-FMR with perpendicular applied magnetic field has been used to determine the gyromagnetic factor. The BLS measurements reveal a pronounced nonreciprocal spin waves propagation, due to the interfacial Dzyaloshinskii-Moriya interaction (DMI) induced by Ir interface with CFA, which increases with decreasing CFA thickness. The DMI sign has been found to be the same (negative) as that of Pt/Co, in contrast to the *ab-initio* calculation on Ir/Co. The thickness dependence of the effective DMI constant shows the existence of two regimes similarly to that of the perpendicular anisotropy constant. The DMI constant $D_s$ was estimated to be -0.37 pJ/m for the thickest samples where a linear thickness dependence of the effective DMI constant has been observed.

**I- Introduction**

The exchange interaction between electrons arises from the Coulomb interaction and is responsible for the microscopic magnetic behavior. This interaction might contain symmetric and asymmetric terms. The symmetric term is commonly known as the Heisenberg [1] interaction (imposing collinear configurations in magnetic structures), while the asymmetric exchange is referred to as the Dzyaloshinskii–Moriya interaction (DMI). For the latter, Dzyaloshinskii [2] predicted, purely on grounds of symmetry, that the combination of low symmetry and spin-orbit couplings gives rise to asymmetric exchange interactions. Moriya found a microscopic mechanism which leads to such a term in systems with spin-orbit coupling [3].

The DMI, which favors canted neighboring spins leading to various magnetization structures at the nanoscale such as helices [4] and skyrmions [5-7], can thus be induced by a lack of inversion symmetry and a strong spin-orbit coupling. Both these requirements are met in heavy metal/ferromagnet (HM/FM) heterostructures, giving rise to the so called interfacial DMI. High values of the MDI constants can be of great value in chiral domain wall (DW) engineering including stabilization of Néel type fast skyrmions, instrumental in implementation of the race-track memory. At the same time, the anti-symmetric nature of the DMI excludes its doubling in a FM layer, sandwiched between two identical HM films; the two contributions are mutually cancelled. However, the situation can be radically changed, if the upper and lower HM films are characterized by IDMI constants with opposite signs, in other words, in an asymmetric configuration.

The DMI is usually characterized by its effective ($D_{eff}$) or surface ($D_s$) constants [8]. It is thus interesting for both application and fundamental research to determine precisely the sign and the value of the DMI constant. Several experimental [9-12] and theoretical studies [13, 14], largely based on how this interaction alters the properties of DW were performed recently. However, the experimental evaluation of $D_{eff}$, using the above mentioned techniques, is at best indirect and based on strong assumptions about the dynamics and magnetization configuration of the DW. Moreover, any numerical estimation is to be checked experimentally: sometimes, discrepancies especially in the DMI sign arise leading to unsuitable sample design [15]. Indeed, recent experiments on asymmetric DW propagation [16, 17] as well as *ab-initio* predictions concluded to opposite DMI signs for Co/Ir and Co/Pt [18]. Chen *et al.* confirmed this sign difference by visualizing the extent of DW chirality in perpendicularly magnetized [Co/Ni]$_n$ multilayers in contact with Pt and Ir [19]. For the above cited studies, the complex structures involving both Ir and Pt or different ferromagnets at the interfaces with Pt and Ir complicate the DMI evaluation for each interface and their comparison. For an unambiguous determination of the DMI sign, Kim et al. [20] investigated experimentally the thickness dependence of the DMI in Ir/Co/AlO$_x$ by means of Brillouin light scattering (BLS) and observed that the Pt/Co and Ir/Co interfaces have the same DMI sign. It should be emphasized that this technique relying on direct measurement of non-reciprocity of spin wave propagation in such structures that scales with the $D_{eff}$ is considered being the most reliable in such studies. Kim *et al.* concluded thus that the DMI energy is quite sensitive to the details of the multilayer structures. Therefore, attention should be paid to the whole stack when concluding about the sign and strength of the DMI constant. This discrepancy raises a debate about the reliability of the *ab-initio* calculations and DW observations in determining the sign of the DMI, which can even be misguiding. For example, in reference [15], authors misguided by these *ab-initio* predictions used asymmetric Ir/Co/Pt multilayers for increasing the effective DMI strength. Therefore, a direct and precise experimental measurement of the DMI sign and magnitude is of outmost importance. Moreover, Co$_2$FeAl is one of the most prominent Co-based Heusler alloys [21] due to its relatively high spin polarization and low magnetic damping parameter [22]. Consequently, in this work we use vibrating sample magnetometry (VSM) and microstrip ferromagnetic resonance combined with BLS to measure magnetization at saturation and gyromagnetic factor for precise analysis of the $Co_2FeAl$ thickness dependence of the DMI constants in $Ir/Co_2FeAl$ ultrathin heterostructures. Our main focus is to address the DMI in such as deposited complex Heusler alloys where, besides their potential application in spintronics, they give the opportunity to investigate the DMI dependence with the atomic distribution at interfaces since their structure and atomic disorder are both annealing and thickness dependent. This latter aspect is very interesting and reveals non regular behavior which can trigger consideration on theories and models to explain the observed trends. The effect of the annealing temperature on the DMI in $Co_2FeAl$ ultrathin will be addressed in forthcoming paper. Moreover, we show that the effective constant demonstrates the pattern of behavior similar to that reported by Kim. *et al*. for Ir/Co [20], it is thickness and interface dependent while its sign is identical to that induced by Pt (in Pt/Co systems).

**II- Samples and experimental techniques**

$Co_2FeAl$ (CFA) thin films were grown at room temperature on a Si substrate covered with a 100 nm thick thermally oxidized $SiO_2$ layer using a magnetron sputtering system with a base pressure lower than $2\times10^{-8}$ Torr. Prior to the deposition of CFA film, a 2 nm thick Ta seed layer and a 4 nm thick Ir layer were deposited on the substrate. Next, the CFA films, with variable thicknesses (0.9 nm≤$t_{CFA}$≤1.8 nm), were deposited at room temperature by dc sputtering under an Argon pressure of 1 mTorr, at a rate of 0.1 nm/s. Finally, in order to protect the structure from air exposure a 2 nm thick Ti film was deposited on top of the CFA layer. In these heterostructures, the Ir layer induces DMI in the CFA ultrathin layers, while Ti is used only to protect CFA from oxidation and is expected to induce no DMI contribution, since it is not a heavy metal.

The crystal structure of the films was studied by x-ray diffraction (XRD) using a four-circle diffractometer. VSM has been used to measure hysteresis loops, both with the magnetic field applied perpendicular and parallel to the films plane, and to determine static magnetic parameters. Microstrip line ferromagnetic resonance (MS-FMR) [22] has been employed here for determining the gyromagnetic factor for the thickest samples ($t_{CFA} \geq 1.2$ nm), for which a MS-FMR signal has been detected.

In the BLS set-up, the spin waves (SW), of a wave number ($k_{sw}$) in the range 0–20 $\mu m^{-1}$ (depending on the incidence angle $\theta_{inc}$: $k_{sw} = \frac{4\pi}{\lambda}\sin(\theta_{inc})$ in backscattering configuration), are probed by illuminating the sample with a laser having a wavelength $\lambda$=532 nm. The magnetic field was applied perpendicular to the incidence plane, which allows for probing spin waves propagating along the in-plane direction perpendicular to the applied field: Damon-Eshbach (DE) geometry where the DMI effect on the SWs propagation non reciprocity is maximal [23]. For each angle of incidence, the spectra were obtained after sufficiently counting photons to have well-defined spectra where the line position can be determined with accuracy better than 0.2 GHz. The Stokes (S, negative frequency shift relative to the incident light as a magnon was created) and anti-Stokes (AS, positive frequency shift relative to the incident light as a magnon was absorbed) frequencies, detected simultaneously were then determined from Lorentzian fits to the BLS spectra. For identical interfaces, S and AS modes should have the same frequency. In the presence of DMI on one interface, the frequency difference between these two propagating SWs exists and increases with $k_{sw}$. Therefore, the DMI constants are determined from $k_{sw}$ dependence of the frequency difference between S and AS lines.

For the analysis of the BLS measurements, the DE mode dispersion [24, 25], taking into account the DMI contribution is given by the equation:

$$F = F_0 + F_{DMI} = \mu_0 \frac{\gamma}{2\pi} \sqrt{(H + Jk_{sw}^2 + P(k_{sw}t_{FM})M_s)(H + Jk_{sw}^2 - P(k_{sw}t_{FM})M_s + M_{eff})} \pm \frac{\gamma}{\pi M_s} D_{eff} k_{sw} \quad (1)$$

here $H$ is the in-plane applied field, $t_{FM}$ is the ferromagnetic layer thickness, $\mu_0$ is the permeability of vacuum and $J = \frac{2A_{ex}}{\mu_0 M_s}$ with $A_{ex}$ is the exchange stiffness constant. The coefficient $P(k_{sw}t_{FM}) = 1 - \frac{1-\exp(-k_{sw}t_{FM})}{k_{sw}t_{FM}}$, describing dipolar interactions, reduces in thin films ($k_{sw}t_{FM} \ll 1$) to a simple $P(k_{sw}t_{FM}) = \frac{k_{sw}t_{FM}}{2}$ which makes this term linear in $k_{sw}t_{FM}$. It should be noticed that in our case $k_{sw}t_{FM} \approx 0.02$.

We define the interfacial DMI constant as $D_s = D_{eff} \times t_{FM}$. From this, the frequency difference can be inferred to be:

$$\Delta F = F_S - F_{AS} = \frac{2\gamma}{\pi M_s} D_{eff} k_{sw} = \frac{2\gamma}{\pi M_s} \frac{D_s}{t_{FM}} k_{sw} \quad (2)$$

According to this equation (1), the dispersion splits into two branches corresponding to the frequency shift in the Stokes $F_S$ and anti-Stokes $F_{AS}$ lines. Each one results from two contributions. While the major one, being field dependent, takes into account the dipole-dipole interactions linear in $k_{sw}$ (in ultra-thin films as ours) and a quadratic in $k_{sw}$ contribution of the conventional isotropic exchange, the DMI contribution, linear in $k_{sw}$, is described by a smaller addition whose sign depends on whether one is interested in the S or AS frequency shift. Importantly, if $F_S$ is lower than $F_{AS}$, then the resulting DMI constant is negative for positive applied magnetic field.

## III- Results and discussion

The measurements presented here were performed at room temperature. Figure 1 shows a 2θ/ω (out-of-plane) x-ray diffraction pattern measured for the sample with $t_{CFA}$ = 1 nm. One can observe that, except for the peak corresponding to the Si substrate, the pattern shows only the (111) Ir peak. This suggest that the Ir layer has a strong (111) out-of-plane texture. The absence of a diffraction peak from the Ta layer indicates, as expected, that the film is in amorphous state. The same result might be valid for the CFA layer, but it is unlikely, having in view that the lower Ir layer has a strong (111) texture. In order to test this, we have grown a sample with a much thicker CFA layer of 6 nm. The inset of figure 1 shows a detail of the 2θ/ω x-ray diffraction patterns of both the 1 nm and 6 nm thick CFA layers samples. A diffraction maximum is clearly visible for the 6 nm thick CFA sample at a 2θ around 44°, which can be attributed to the (022) CFA reflection. The absence of the (022) diffraction peak for the 1 nm thick CFA films is a consequence of the ultra-low thickness of the film corroborated with the relative low atomic scattering factors of the CFA constituents. No other additional diffraction peaks were observed for the 6 nm thick CFA sample as compared to the 1nm thick CFA samples. This indicates that the CFA films shows a (022) out-of-plane texture. Furthermore, φ-scan measurements (not shown here) showed that both Ir and CFA have no in-plane texturing but in-plane isotropic distribution of the crystallites.

Figures 2(a) and (b) show in-plane and out-of-plane hysteresis loops measured for the sample with a CFA thickness of 1.8 nm. The out-of-plane hysteresis loop indicates a continuous rotation of the magnetization towards the perpendicular direction as the magnetic field is increased. This indicates that the sample possess an in-plane anisotropy easy axis. A weak uniaxial anisotropy was observed in the plane, as indicated by the different shape of the hysteresis loops measured in-plane [Fig. 2(a)]. The presence of small uniaxial in-plane anisotropy is not unusual for sputtered films and it is due to a residual magnetic field present during growth. It is to be mentioned that the other samples show a similar behavior, except the sample with a CFA thickness of 0.9 nm, whose in-plane and out-of-plane hysteresis loops are shown in Fig. 2(c) and (d). As we will see below, this sample is at the limit between in-plane and perpendicular magnetic anisotropy and most likely it shows complex domain structure rendering the null remanence magnetization. We should mention that the magnetization in figure 2 was evaluated by considering the nominal CFA thickness, which explain the difference of the magnetization at saturation between the 1.8 nm and 0.9 nm thick samples since the magnetic dead layer is not taken into account.

Figure 3 (a) depicts the CFA thickness dependence of the saturation magnetic moment per unit area, which is used to determine the magnetization at saturation ($M_s$) and the magnetic dead layer thickness ($t_d$), as the slope of the linear fit of the data gives $M_s$, while the horizontal axis intercept gives $t_d$. The thickness of the magnetic dead layer and magnetization at saturation are found to be 0.44 nm and 1035±55 emu/cm$^3$ (error bar less than 6%). The magnetic dead layer is most probably due to intermixing at the Ir/ CFA interface, since the deposition of heavy metal onto ferromagnet (or vice versa) is usually accompanied by such mixing effects. Therefore, the magnetic dead layer should be taken into account for the CFA effective thickness to be used when determining the effective anisotropy and DMI constants. Even though a dead layer at the bottom interface exists, this does not completely cancel the DMI interaction, as it will be experimentally shown bellow. The increase of $M_s$ for the Ir/CFA system, compared to that of MgO/CFA/MgO ($M_s$~850±50 emu/cm$^3$) [26] is most likely due to the proximity induced magnetization in Ir. This corresponds to a change in film magnetization of 22%, which is slightly higher than the reported value (19%) in Ir/Co/Ni/Co [27] and Ir/Co [20] systems.

The *g* value, which determines the gyromagnetic factor and therefore the precision on the evaluation of the DMI constant, is precisely accessible by the MS-FMR technique using, through the study of the frequency variation versus the magnetic field applied perpendicularly to the film plane. Typical MS-FMR perpendicular field dependence of the resonance frequency is shown in Fig. 3(b). The linear variation as function of the magnetic field is in agreement with the expected theoretical dependence $F_{\perp} = \left(\frac{\gamma}{2\pi}\right)(H - 4\pi M_{eff})$, where $(\gamma/2\pi) = g \times 1.397 \times 10^6$ Hz/Oe is the gyromagnetic factor and $M_{eff}$ is the effective magnetization [22]. The derived value of $g$ = 2.04 ($\gamma/2\pi$=29.2 GHz/T) is in excellent agreement with the value determined in our previous papers [22, 28] for relatively thick CFA films. Since this value does not present a significant variation versus the CFA thickness (at least for the thickest CFA films ($t_{CFA} \geq 1.2$ nm), for which a MS-FMR signal has been detected), it will be used for all the samples studied here.

To quantify the magnitude of the out-of-plane magnetic anisotropy of our films we determined the effective perpendicular magnetic anisotropy constant $K_{eff}$ from the saturation field ($H_s$), using the relation $K_{eff} = -M_s H_s / 2$. The $H_s$ value was determined from the perpendicular out-of-plane hysteresis loops [see Fig. 2(b)]. Phenomenologically, the $K_{eff}$ dependence on thickness can be separated into a volume ($K_v$) and a surface contribution ($K_s$) as $K_{eff} \times t_{eff} = K_v \times t_{eff} + K_s$, where $t_{eff} = t_{CFA}$-$t_d$ is the CFA layer effective thickness [29,30]. As depicted in Fig. 4(a), the $K_{eff} \times t_{eff}$ does not show a single linear dependence on the $t_{eff}$ for the whole thickness range. Instead, there are two regions separated by a critical thickness $t_c$, each with its own linear dependence, characterized by different slopes. Several explanations can be given to the deviation from the single linear behavior of the perpendicular effective anisotropy versus the CFA effective thickness [29] and therefore to the existence of a second regime of higher effective anisotropy, as shown in figure 4. The four most important mechanisms will be discussed separately and will be used to analyze our experimental data. Firstly, a possible coherent–incoherent growth transition, with the accompanying changes in the magneto-elastic anisotropy contributions can lead to this two regimes behavior, commonly observed in thin films systems in which there is an elastic strain relaxation above a certain critical thickness [29-30]. In the case of our samples, since Ir and CFA grows with a (111) and (011) out-of-plane texture and having in view the lattice parameters of the two films, we expect that in the first stages of growth CFA to be subdued to an in-plane compressive stress which at least partially relaxes through the formation of misfit dislocations as thickness is increased.

In order to analyze the results and according to the model from [29, 30], we will consider two regimes bellow (regime I) and above (regime II) the critical thickness, in which $K_v$ and $K_s$ are given by:

$$\begin{cases} K_v^I = -2\pi M_s^2 + K_{mc} + K_{me,v} \\ K_s^I = K_N \end{cases} \quad \text{in regime I} \tag{3}$$

$$\begin{cases} K_v^{II} = -2\pi M_s^2 + K_{mc} \\ K_s^{II} = K_N + K_{me,s} \end{cases} \quad \text{in regime II} \tag{4}$$

Here $K_{mc}$ is the magnetocrystalline anisotropy, $K_{me,v}$, $K_{me,s}$ are the volume and interface magneto-elastic anisotropy constants, $2\pi M_s^2$ is the shape anisotropy contribution and $K_N$ is Néel-type perpendicular interface anisotropy constant induced by the broken symmetry at the interfaces. According to this model, in region I, the influence of misfit strain appears as a volume contribution (characterized by $K_{me,v}$) to the anisotropy and, although it is bulk related, it leads to an apparent interface contribution in regime II [29, 30] (characterized by $K_{me,s}$).

The linear fit of the measurements of Fig. 4 allows us to determine constants for both regimes from the slope and the intercept with the vertical axis, respectively. Then by using equations (3) and (4), the contributions of the magneto-crystalline, magneto-elastic and the Néel-type interface anisotropies to the surface and volume perpendicular anisotropies have been isolated: $K_{mc}$= (2.6±0.1)×10$^6$ erg/cm$^3$, $K_{me,v}$= -(2.2±0.6)×10$^6$ erg/cm$^3$, $K_{me,s}$ = -(0.18±0.05) erg/cm$^2$ and $K_N$ = (0.32±0.03) erg/cm$^2$. The magnetoelastic anisotropy is negative reinforcing the in-plane easy axis. The Néel-type surface interface anisotropy, reinforcing perpendicular easy axis, can be attributed to the Ir/CFA interface [30]. Both volume and surface magnetoelastic anisotropy are negative and thus reinforce the in-plane easy axis. This is coherent with the fact that CFA films are in-plane compressed and with the positive magnetostriction coefficient of CFA [32]. In order to furthermore confirm the observed trend of the out-of-plane magnetic anisotropy of our films, we have determined the effective magnetization ($4\pi M_{eff} = 4\pi M_s - \frac{2K_\perp}{M_s}$, where $K_\perp$ is the perpendicular anisotropy constant) using both BLS [for the thinner samples (0.9 nm≤$t_{CFA}$≤1.1 nm), where the MS-FMR signal was not sufficient to follow the field dependence of precession frequency] and MS-FMR techniques. The extracted values are shown in figure 4b, as function of 1/$t_{eff}$. Depending on $t_{CFA}$, two different regimes, separated by a critical thickness $t_c$ (nominal CFA thickness around 1 nm) can be distinguished. For both regimes, $M_{eff}$ decreases linearly with 1/$t_{eff}$ but with different slopes: the slope is higher for $t<t_c$. The linear fit of the measurements of figure 4b allows determining the perpendicular surface and volume anisotropy constants for both regimes from the slope and the intercept with the vertical axis, respectively, since the perpendicular anisotropy constant $K_\perp$ obeys to the relation $K_\perp = K_{v\perp} + \frac{K_s}{t}$. Then by using equations (3) and (4), the MS-FMR anisotropy constants ($K_{mc}$= 1.84×10$^6$ erg/cm$^3$, $K_{me,v}$= - $3.04\times10^6$ erg/cm$^3$, $K_{me,s}$ = -0.21 erg/cm$^2$ and $K_N$=0.384 erg/cm$^2$) are in good agreement with the ones deduced from the static measurements.

Another possible way to explain the two regimes behavior is the roughness in the thinner films. Such roughness creates in-plane demagnetizing fields at the edges of terraces reducing the shape anisotropy and therefore, favors perpendicular magnetization: the effective magnetization $4\pi M_{eff}$ is modified into $4\pi M_{eff} = (4\pi - N_x - N_y)M_s - \frac{2K_\perp}{M_s}$ where $N_x$, $N_y$ are the in-plane demagnetizing factors. In the case of a perfectly flat film, $N_x = N_y = 0$, $N_z = 4\pi$, while edges of discontinuities yield an increase of $N_x$ and $N_y$. The influence of the roughness has been calculated in frame of dipolar approximation by H. Szymczak *et al.* [33]: $(4\pi - N_x - N_y) = 4\pi - 3\pi(\sigma/t)(1-f)$, where $\sigma$, which is a statistical parameter characterizing roughness, is the average deviation from the reference plane and $f$ is a tabulated factor depending on the geometric parameters. According to this model, the surface anisotropy constant due to the roughness ($K_r$) is given by:

$$K_r = \frac{3}{2}\pi M_s^2 \sigma(1-f) \tag{5}$$

In the regime of thinner films, the surface anisotropy constant is thus $K_s^I = K_s^{II} + K_r$. By using the values $K_s^I$ and $K_s^{II}$ obtained from the linear fit of the experimental data shown in figure 4, we determine $\sigma\sim0.6$ nm and $f\sim0.2$ (figure 4 in ref. [33]). This is very high roughness value is not reasonable since the usual measured one in such samples is about 0.3 nm. Moreover, as the effective CFA thickness is comparable to $2\sigma$ (terrace height) in the case of thinner films, this roughness value implies the occurrence of discontinuities in the thinner films. Moreover discontinuities in the CFA films yield a lower effective magnetic/non-magnetic interface area, thus a lower interface contribution and a correspondingly lower total anisotropy and consequently an increase of the effective magnetization. Therefore, to take the discontinuities effect on interface anisotropy into account, one should consider a roughness larger than the above estimation σ > 0.6 nm yielding a terrace height superior to 1.2 nm which is not meaningful because the thinnest film thickness is inferior to this value.

Finally, interdiffusion and mixing might occur at the interfaces during the deposition of the layers introducing thus, randomness in the magnetic pair bonds according, which obviously reduces the interface anisotropy [29]. This latter mechanism is incompatible with the experimental results shown in figure 4, where a higher effective anisotropy is observed for thinner films: below $t_c$. The consistence of the two first models with our experimental results will be further discussed below after presenting the determination of the DMI constant.

Figure 5 shows the typical BLS spectra for the 1.4 nm thick sample for $k_{sw}$=18.1 μm$^{-1}$ ($\theta_{inc}$=50°) and 20.45 μm$^{-1}$ ($\theta_{inc}$=60°). It reveals the existence of both S and AS spectral lines. Besides the usual intensity asymmetry of these lines due to the coupling mechanism between the light and SWs, a pronounced difference between the frequencies of the S and AS modes ($\Delta F=F_S-F_{AS}$), especially for higher values of $k_{sw}$, is revealed by the BLS spectra. This frequency mismatch is due to the interfacial DMI as demonstrated previously [8, 23, 25]. Since the inverse proportionality on the ferromagnetic layer thickness is usually a signature of an interface effect, the behavior of $\Delta F$ versus $1/t_{eff}$ is presented in the insert of Fig. 6a for $k_{sw}$=20.45 μm$^{-1}$ ($\theta_{inc}$=60°). It can be observed that $\Delta F$ increases with $1/t_{eff}$ and approaches zero when $t_{CFA}$ tends to infinity confirming the interfacial origin of the DMI. Figure 6a shows the $k_{sw}$ dependence of $\Delta F$ for CFA thin films of various thicknesses, where a clear linear behavior can be observed. From the slopes of the $k_{sw}$ dependences of $\Delta F$, the effective DMI constants have been extracted using equation (2) with $\gamma/(2\pi)$=29.2 GHz/T and $M_s$=1035 emu/cm$^3$ deduced from the fit of MS-FMR data and the VSM measurements, respectively. The evolution of the obtained values of $D_{eff}$ as function of the inverse of the CFA films effective thickness ($1/t_{eff}$) are shown in Fig. 6b where a linear behavior can be observed, as predicted theoretically. Note the deviation from the linearity, as the CFA nominal thickness approaches to 1.1 nm similarly to the thickness dependence of the perpendicular anisotropy (Fig. 4): two regimes (above and below CFA nominal thickness of 1.1 nm) with different slopes can be distinguished. By the linear fit of the data of figure 5b for $t_{CoFe} \geq 1.1$nm, $D_s$ has been found to be –0.34 pJ/m. This value is significantly lower than that of Pt/Co/AlO$_x$ systems [17] but has the same sign as Pt/Co confirming the recent results of Kim et al. [20] for Ir/Co. However, it is slightly lower than the one measured for Ir/Co (-0.8 pJ/m) [20] most probably due to the fact that the authors have ignored the magnetic dead layer when determining $D_s$. Moreover, CFA films (as all the Heusler alloys) are subject to some degree of chemical disorder, which strongly influences many of their physical properties and thus DMI. Thinner films (thickness below 10 nm) annealed at low temperature (below 300°C) have the A2 structure, corresponding to a complete disorder between all atoms Co, Fe, and Al [26]. Therefore, within the CFA thickness range presented in this paper, all films have mostly the same disordered A2 structure, which may explain the smaller DMI constant in Ir/CFA compared to Ir/Co. By considering the evolution of the obtained values of $D_{eff}$ as function of the inverse of the CFA films nominal thickness (not shown here) the deduced value of $D_s$ (-0.51pJ/m) is comparable to that of Ir/Co [20]. We should mention that the existence of the two regimes of the thickness dependence of $D_{eff}$ has been observed for Pt/CoFe systems [34] and Pt/CoFeB [35], in contrast to Pt/Co systems [35]. Moreover, this piecewise linear behavior seems to be a characteristic of alloyed ferromagnetic films. It seems being more significant as the number of atoms constituting the alloy increases. For example, the two regimes of the thickness dependence of $D_{eff}$ have different slopes with the same sign in the case of CoFe [34] while an inversion of the trend has been observed for CoFeB [35] and here for CFA. Although the diminution of $D_{eff}$ as thickness decreases was directly correlated to interface degradation in the case of Pt/CoFe [34] of Pt/CoFeB [35], this correlation is not obvious in Ir/CFA since the slopes of the thickness dependence of $M_{eff}$ and $D_{eff}$ for ultrathin films ($t_{eff}<t_c$) have opposite signs: for $t_{eff}<t_c$, the effective surface anisotropy (surface DMI constant) is higher (lower) than that for $t_{eff}>t_c$. However, the existence of the two regimes can be understood through the above mentioned first and second mechanisms for the perpendicular anisotropy. Within the optics of the first mechanism, for the thickest CFA films ($t_{eff}>t_c$), the growth induced stress is relaxed by dislocations at the interface, and $D_s$ and $D_{eff}$ are defined by $D_{eff} = D_s/t_{eff}$, as observed in figure 6b. In the regime of low thicknesses ($t_{eff}<t_c$), the CFA films are strained (compressive stress) and the interface is without dislocations. Therefore, distance between Ir and CFA atoms at the interface changes, modifying thus the DMI constant, according to Fert et al. [36]. In the frame of the second approach mentioned above, CFA films discontinuities at interfaces decrease the contact surface between Ir and CFA reducing thus both the DMI and the anisotropy constant for thinner. Finally, it is worth to mention that Nembach et al.[37] demonstrated a linear proportionality between the exchange stiffness constant and DMI. Furthermore, they observed a non-linear thickness dependence of the effective DMI constant in Pt/Py and speculated that this non-linear behaviour is the result of an unexpected thickness dependence of the symmetric exchange for this particular system. The microscopic origins of the variation of the symmetric exchange with film thickness are unclear and it is an empirical fact that both the symmetric exchange and the asymmetric exchange exhibit the same nontrivial functional dependence on reciprocal thickness as acknowledged by. Nembach [37]. Since it is not possible to precisely measure the exchange stiffness constant by BLS for such ultrathin CFA films the thickness dependence of the exchange stiffness constant in our ultrathin films cannot be checked and thus it is not possible to speculate about this behaviour. However, this possible interpretation of the thickness dependence of the DMI cannot be verified.

Due to the lack of precise information about residual strain, interface quality and thickness dependence of the exchange stiffness constant, it is not obvious to identify the mechanism responsible for both the decrease of $D_{eff}$ and the increase of the surface anisotropy for the thinner CFA films. However, we strongly believe that the CFA film discontinuities at the interface with Ir are not compatible with thickness dependence of the effective anisotropy and therefore, the probably responsible mechanism for both behaviours of DMI and effective perpendicular anisotropy is possible coherent–incoherent transition.

**Conclusions**

CFA films of various thicknesses (0.9 nm≤$t_{CFA}$≤1.8 nm) were prepared by sputtering on Ta/Ir-buffered Si/SiO$_2$ substrates. The vibrating samples magnetometry measurements revealed that the CFA films are in-plane magnetized. Ferromagnetic resonance with a microstrip line has been used to determine the gyromagnetic factor and Brillouin light scattering has been employed in the Damon-Eshbach geometry to investigate the spin waves non reciprocity induced by the interfacial Dzyaloshinskii-Moriya interaction (DMI). It turned out that the DMI effective constant sign of Ir/CFA is the same as the Pt/Co and Ir/Co ones (deduced from BLS experiments), in contrast to that of Ir/Co which was found to be of opposite sign according to both the theoretical calculations and some experimental observations on such Ir/Co systems. Indeed, recent experiments on asymmetric DW propagation as well as *ab-initio* predictions both point to opposite DMI signs for Ir/Co and Pt/Co.

**Acknowledgements**

M.B would like to thank A. Thiaville for fruitful discussions and remarks. This work has been partially supported by the Conseil Régional, Île-de-France through the DIM C'Nano IMADYN project. M.S.G., T.P. and R.B.M. acknowledge the financial support of UEFISCDI through PN-II-RU-TE-2014-4-1820 SPINCOD Research Grant No. 255/01.10.2015.

**Fig. 1 : Belmeguenai et al.**

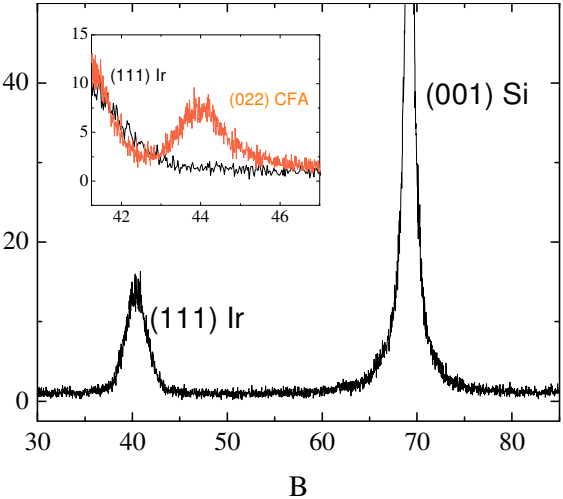

**Fig. 2 : Belmeguenai et al.**

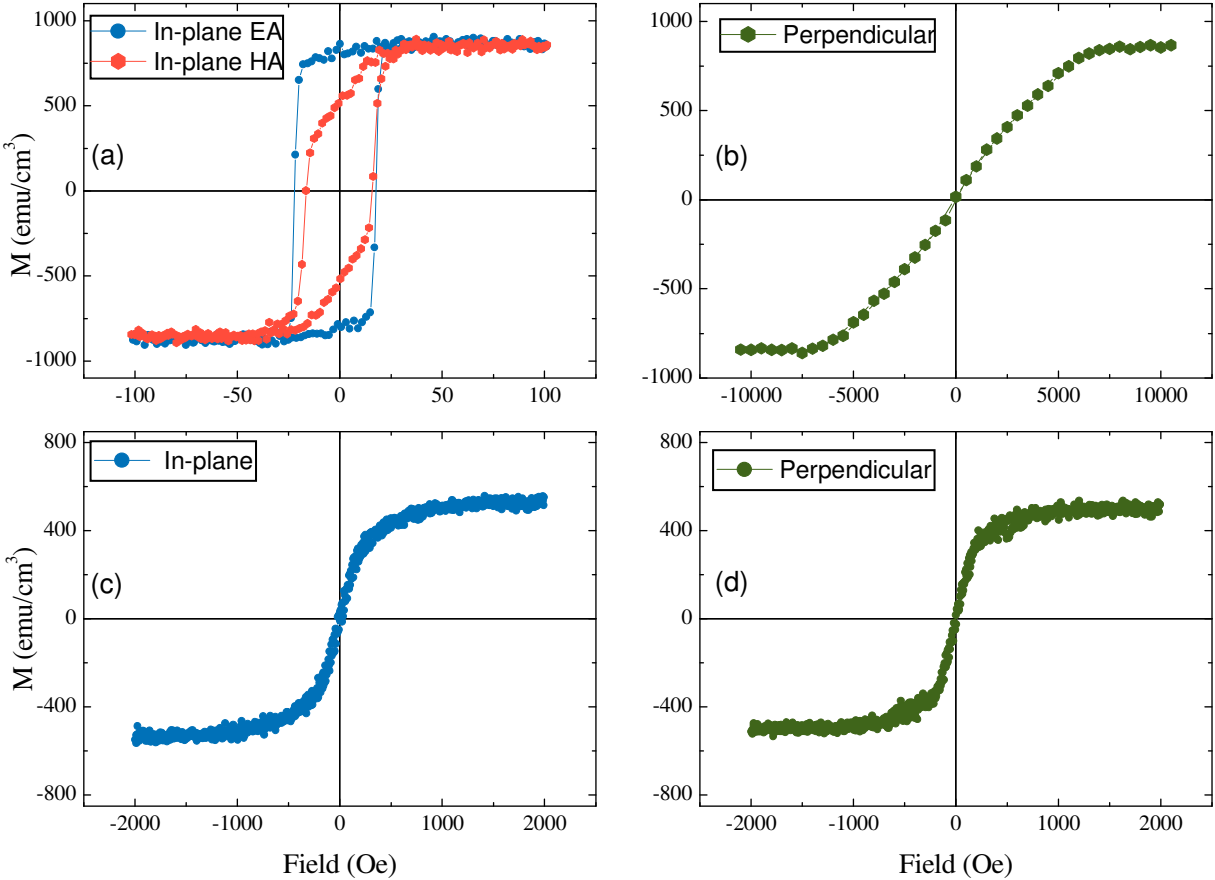

**Fig. 3 : Belmeguenai et al.**

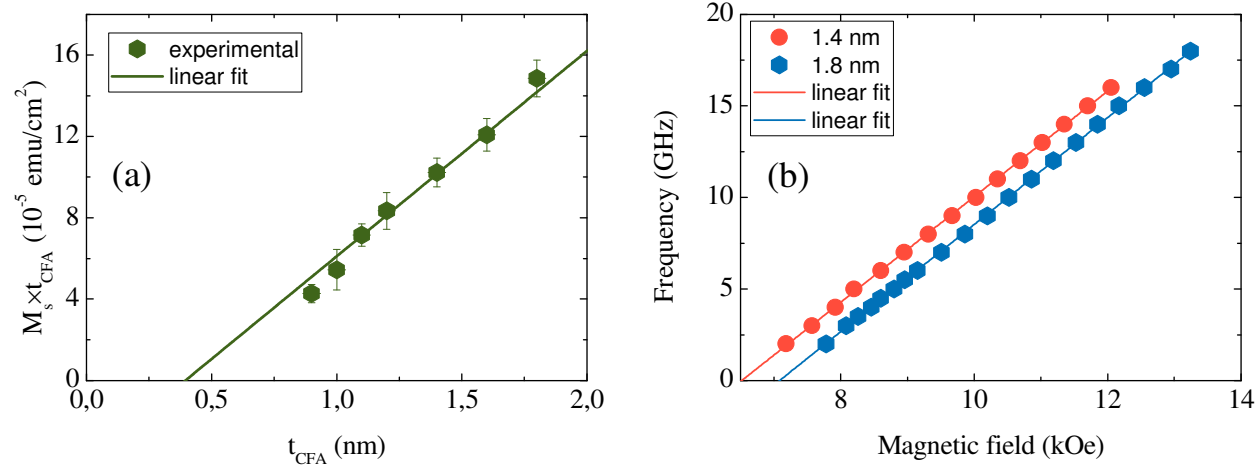

**Fig. 4 : Belmeguenai et al.**

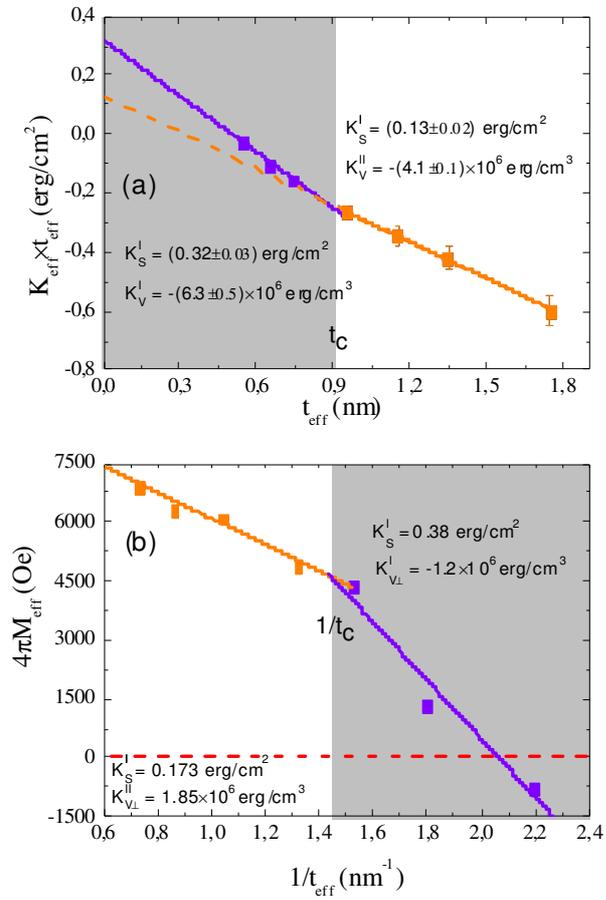

**Fig. 5 : Belmeguenai et al.**

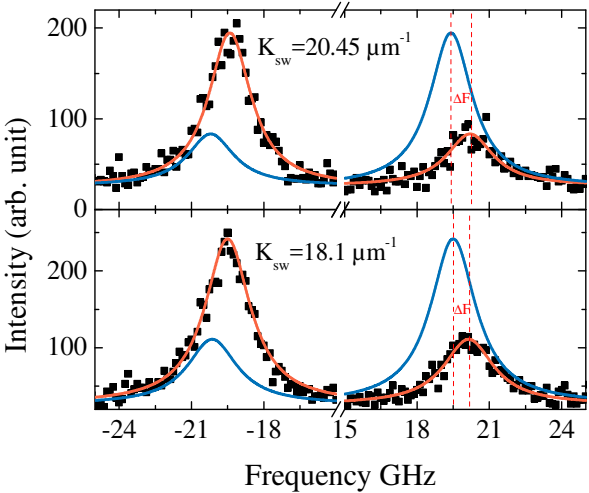

**Fig. 6 : Belmeguenai et al.**

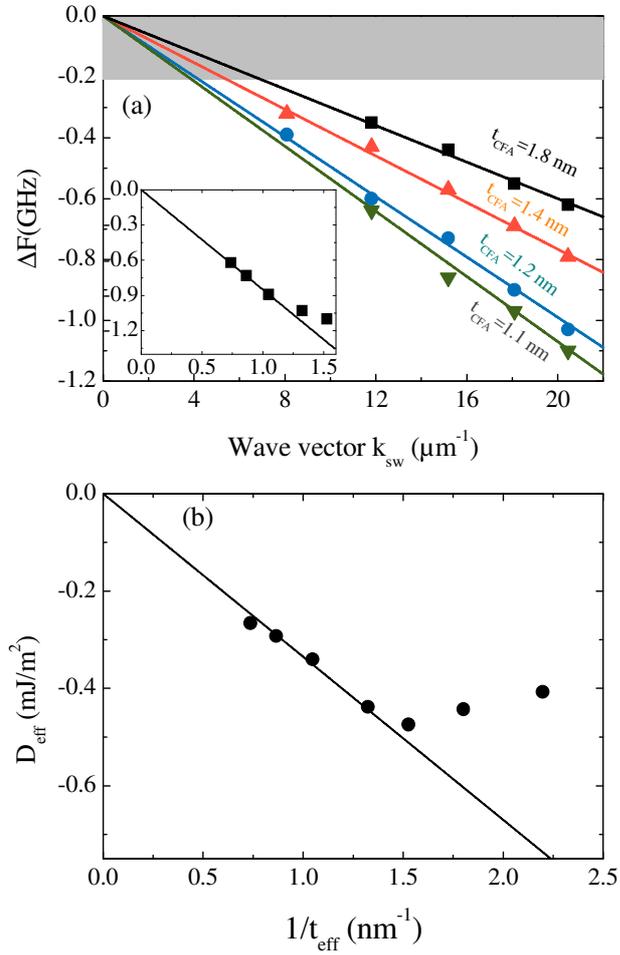

**Figure captions**

**Figure 1:** (Color online) 2θ/ω (out-of-plane) x-ray diffraction pattern measured for the sample with $t_{CFA}$ = 1 nm. The inset shows a detail of the 2θ/ω x-ray diffraction patterns of both the 1 nm and 6 nm thick CFA layers samples.

**Figure 2:** (Color online) In-plane (a) and out-of-plane (b) hysteresis loops measured for the sample with a CFA thickness of 1.8 nm. In-plane (c) and out-of-plane (d) hysteresis loops measured for the sample with a CFA thickness of 0.9 nm. The magnetization in figure 2 was evaluated by considering the nominal CFA thickness

**Figure 3:** (Color online) (a) The thickness dependence of the of the saturation magnetic moment per unit area. (b) MS-FMR perpendicular field dependence of the resonance frequency measured for the samples with a CFA thickness of 1.4 nm and 1.8 nm, respectively.

**Figure 4:** (Color online) (a) $K_{eff} \times t_{eff}$ versus effective CFA thickness deduced from perpendicular applied magnetic field hysteresis loops. (b) Thickness dependence of the effective magnetization ($4\pi M_{eff}$) extracted from the fit of BLS ($t_{CFA}$<1.2 nm) and MS-FMR ($t_{CFA} \geq$1.2 nm) measurements Symbols refer to experimental data while solid lines are the linear fits corresponding to two regimes. For direct comparison between the anisotropy constants indicated in figures (a) and (b), note that $K^i_v = K^i_{v\perp} - 2\pi M_s^2$, where the superscript $i$ refers to regime I or II.

**Figure 5:** (Color online) BLS spectra measured for 1.4 nm thick CFA film at 4 kOe in-plane applied magnetic field values and at two characteristic light incidence angles corresponding to $k_{sw}$ = 18.1 and 20.45 $\mu m^{-1}$. Symbols refer to the experimental data and solid lines are the Lorentzian

fits. Fits corresponding to negative applied fields (blue lines) are presented for clarity and direct comparison of the Stockes and anti-Stockes frequencies.

**Figure 6:** (Color online) (a) Wave vector ($k_{sw}$) dependence of the experimental frequency difference $\Delta F$ of CFA films of a thickness $t_{CFA}$ grown on Si substrates. Solid lines refer to linear fit using Eq. (2) and magnetic parameter in the text. The insert of the figure shows the frequency difference $\Delta F$ of the CFA films for light incidence angles corresponding to $k_{sw}$ = 20.45 as function of the inverse of the effective thickness of CFA films (1/$t_{eff}$). Symbols refer to experimental data and straight solid line is the linear fit. (b) Thickness dependence of the effective DMI constants extracted from fits of Fig. 5a. Solid lines refer to the linear fit.